# Intrusions Detection System Based on Ubiquitous Network Nodes

Lynda Sellami
Department of Computer Science
Bejaia University, Algeria
slynda1@yahoo.fr

Djilali Idoughi
Laboratory of Applied Mathematics
Bejaia University, Algeria
djilali.idoughi@gmail.com

Abderrahmane Baadache
Laboratory of Modeling and Optimization Systems
Bejaia University, Algeria
abderrahmane.baadache@gmail.com

*Abstract*—Ubiquitous computing allows to make data and services within the reach of users anytime and anywhere. This makes ubiquitous networks vulnerable to attacks coming from either inside or outside the network. To ensure and enhance networks security, several solutions have been implemented. These solutions are inefficient and/or incomplete. Solving these challenges in security with new requirement of Ubicomp, could provide a potential future for such systems towards better mobility and higher confidence level of end-user services. We investigate the possibility to detect network intrusions, based on security nodes abilities. Specifically, we show how authentication can help build user profiles in each network node. Authentication is based on permissions and restrictions to access to information/services on ubiquitous network. As a result, our idea realizes a protection of nodes and assures security of network.

## I. INTRODUCTION

The goal of the current research in distributed computing is to abstract the physical location of users and remote resources [1][2]. Such an environment is exposed to serious security threats that can reach people and equipment, as well as data and programs in the virtual world. Therefore, the traditional mechanisms that focus only on digital security become inefficient. It is important to detect security breaches when they occur. This is made possible by intrusion detection mechanism.

Intrusion Detection Systems (IDS) allow to quickly implement new security policies to detect and react as quickly as possible against attacks occurring in a network [3].

IDS have already been used in classic and traditional environments for overcoming intrusion problems [3]. Our goal is to present some IDSs that have been developed for ubiquitous environments and examine their limits in different areas of Ubicomps.

In this article, we propose an IDS that aims to overcome the problems of intrusions in ubiquitous environments. We explore our approach for detecting intrusions (attacks) that occur in ubiquitous environments.

The rest of the paper is organized as follows: Section II presents a state of work already done on intrusion detection in ubiquitous environments. In Section III details our proposal. Finally, conclusion and perspectives are described in Section IV.

## II. RELATED WORK

The use of networks and information systems as tools becomes necessary for the proper functioning and growth of enterprises. The multitude of network usage by known or unknown persons (users), turns the networks into potential targets for attacks. Users can exploit the vulnerabilities of networks and computer systems to access information or to undermine their good functioning.

The security of these networks, targeted by attackers, is an issue of paramount importance. For this, IDSs have been widely discussed for solving the problems of intrusions in networks. Several solutions that have been adopted in order to overcome the problems of attacks.

### A. Ubiquitous Environments

An Intrusion Detection System can be considered as an application, in which individuals and organizations often express the need and objective of protecting their systems against intrusions.

To face the problem of security and the new requirements of the Ubicomp, we introduce and describe the main research work in this direction.

*1) SUIDS (for Service-oriented and User-centric Intrusion Detection System):* This is an IDS for ubiquitous environments; it is a system for intrusion detection of oriented services. SUIDS is suitable for users and deals with smart homes and offices [4][5]. It adopts a new mechanism-oriented service geared to verify and represent user and protect a variety of devices in the network against intrusions.

SUIDS treats the issue of heterogeneity of pervasive networks in three categories, which are principal nodes, service nodes and user nodes. SUIDS organizes the system hierarchically in three levels. The first level corresponds to the domain manager node. The second level contains the principal nodes. The third level is dedicated to service users nodes. To detect intrusions, SUIDS builds on the events of user behavior.

The detection algorithm uses a chi-square test to determine abnormalities. The chi-square test is applied for each specific parameter.

Figure 1 presents Hierarchical System for Smart-Home:

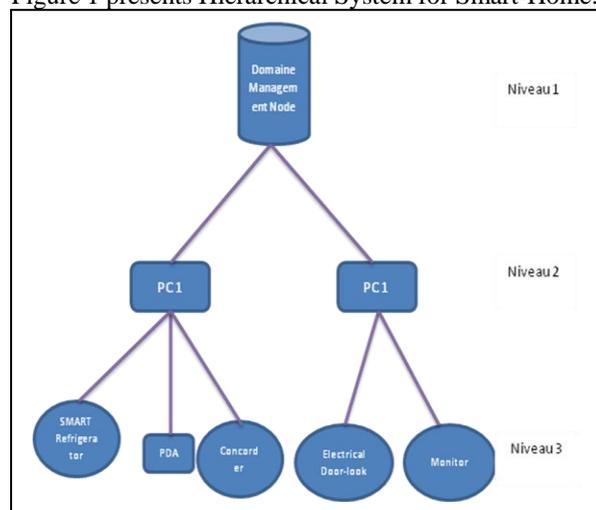

Figure 1. Hierarchical System for Smart-Home [4].

*2) IDS for IP-Based ubiquitous sensor networks:* RIDES (for Robust Intrusion Detection System for IP-Based Ubiquitous Sensor Networks) [6] is a hybrid system of intrusion detection. It combines the signatures of attacks and the discovery of anomalies (behaviors) for detecting intrusions. It uses a distributed algorithm of pattern matching for the intrusion detection based on signatures. To ensure the detection of intrusion based anomalies, RIDES employs a technique of classification based on the notation of SPC (Statistical Process Control) technique called CUSUM chart.

*3) An application-oriented solution:* That enforcement-led approach to security is presented by Robinson [7], as a background of research related to security in ubiquitous computing. It offers a good balance between theory, technology and scenarios. It is a method of generalized research-oriented application and it is applied on a thesis about intrusion detection. This approach is composed of four (4) steps. Step 1 is responsible for identifying the scope and objectives to be achieved. Step 2 provides the design strategy for achieving the objectives of the application. Step 3 provides hardware and software which can be extended or establishes mechanisms to achieve the objectives. Step 4 evaluates theory and technology proposals based on objectives and identified constraints.

*4) Intrusion detection for wireless sensor networks based cluster:* The objective of this approach is to detect and prevent intrusions in sensor networks [8]. Singh et al. [2] have implemented the MAC (Media Access Control), address of intruders attacking networks.

The sensor network is composed of a static Base Station (BS) and clusters. The BS is located away from sensors, and the clusters are composed of a number of dynamic sensor nodes.

Each cluster leader collects data, compresses and transmits them to the Base Station [2]. The Base Station keeps track of the healthy state of all nodes in each cluster by checking information sent by each cluster head to MAC address [2].

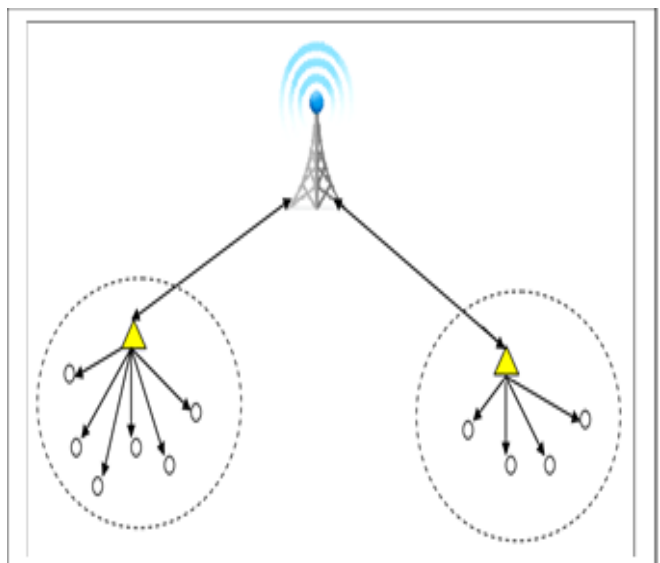

Figure 2. Data communication between base station and heads of clusters [2].

*5) Distributed PCA (Principal Components Analysis) approach for intrusion detection in sensor networks:* Loo et al. [9] present a new approach for distributed intrusion detection called PCADID, for attacks detection in wireless sensor networks.

The sensor array is partitioned into groups of sensor nodes. In each group, a selected control node cooperates with other control nodes in order to build a normal comprehensive profile of the network [9].

The authors have used two (2) approaches [9]:
- An approach based on the PCA centralized approach, called PCACID. Here each control node establishes a normal profile of its own network traffic using PCA (Principal Components Analysis).
- An approach based on the PCA distributed approach, called PCADID. Here each control node using the APC establishes its own normal sub-profile for its traffic network, and sends it to other control nodes. Each control node constructs the normal profile of the network

*6) Policy of intrusion detection for ubiquitous sensors networks:* Xu et al. [10] propose an intrusion detection policy for ubiquitous sensors networks. Their approach consists of three (3) units:
- Unit of data collection: Each sensor node listens to communications among neighboring nodes. The streamed packets are transmitted to the processing unit.
- Unit of data processing: The header of packets received by the playback unit is interpreted and analyzed; the values are then updated in the list of verification data.
- Handling policy: Compare the current activity of the sensors node with the threshold values. If the behavior violates these values, the node is identified as compromised.

## B. Limits and New Motivations

The majority of IDSs developed for ubiquitous environments are designed for specific areas of application or for application scenarios, which limit their generalizations for all areas of Ubicomp. We deal with the following problems already presented in field of mobile ad hoc networks [11] and wireless network [12]: (1) centralization of detection, (2) transition to SMS communications, (3) excessive consumption of energy, (4) capacity limitation of sensors (hardware), and (5) reconfiguration problem when changing the architecture and the components. The IDSs developed for Ubicomp are insufficient or incomplete, and they need improvements and/or adaptation.

TABLE I. IDSs LIMITS OF UBICOMPS

| IDS | Inconveniences |
| --- | --- |
| SUIDS | -The principal nodes are far from the service nodes. They need more energy<br>-The network connections are consumed in communication with service nodes<br>-Intrusions that occur during transmission may go unnoticed.<br>-Existence of the concept of centralized detection can not be applied to larger areas (administrative building).<br>-This solution is not applicable to other areas of applications of ubiquitous where we need a mass of information to convey and treat<br>-Risk of false positives that generates too alarms and treatments. |
| Application-led | -This document provides a general methodology to search applications oriented in Ubicomp.<br>-It is applied on application scenarios<br>-The solution may not work on all instances of an application or other applications with different requirements.<br>-The notion of reconfiguration and adaptation must be added. |
| Cluster-based | - It is based on a central controller |
| ACP-based | -It is necessary that each control node up-to-date has normal profile for PCACID,<br>-It is necessary that each control node up-to-date sounds under normal profile and cooperate with other control nodes to update the overall normal profile of the network.<br>-The normal behavior of the network (sensors) changes over time |
| Policy detection | -Overload sensor nodes through the collection, analysis, transmission and detection. |

## III. PROPOSITION

In this research, we propose an IDS for identifying intrusion from legitimate users in ubiquitous systems.

### A. Description of the Proposal

The ubiquitous network consists of several nodes (devices, hosts, sensors, etc.). It is difficult or impossible to have a global view of ubiquitous network since they are on very large scale. For supporting the scalability of ubiquitous networks, we focus on network nodes. Each node has knowledge of closest neighbors surrounding it. Considering a node and its neighbors allow to extend our approach to a large number of nodes.

Therefore, each treatment applied to a node will be applied to its neighbors.

It is assumed that each node of the ubiquitous network is provided with the collection unit, and the detection and analysis unit. In the case of adhesion of a new node to the network, its neighbors send him the two units after an existence test.

Each node of the network has permissions and privileges, allowing to perform a number of processing and communication with other nodes on the network. Nodes use an adapted routing protocol so that they can route messages among them.

### B. Overview of the Proposition

Figure 3 illustrates the functioning of our proposal; it consists of four parts, namely, data source, behavior, control, and action.
Data flow is defined by the node (sensor, network packets, users query, etc.). The behavior of the node is constructed from the data stream. To detect anomalies in the behavior of nodes (users), a unit of control and analysis is used to compare the actual behavior with the normal profile. When an abnormality is detected, an alarm is sent to the node to help solve this problem. An abnormality corresponds behavior which derived from normal (deviation), or unintended (unexpected).

*1) Detection approach:* There are two ways of detecting intrusions, (1) through known attack patterns to match with, misuse detection, and (2) through expected normal behaviors and identifying deviations as intrusion, anomaly detection. In our case, we have used the anomaly detection technique for identifying intrusions. This, obviously, requires defining the expected normal behaviors of the user that use the service of network. Each node creates its normal profile.

*2) Normal profile:* Each user accesses a ubiquitous network with authentication, which limits access to information/services offered by the network. Authentication uses the notions of permissions and restrictions of the network. This permissions and restrictions allow to build normal profiles of users.

Each node collects information to determine the vector of restrictions, and permissions for network access. These feature vectors (restrictions and permissions) are the normal traffic profile of nodes. Let $V_{i,k}(0)$ be the feature vector of node i of size n, that represents the number of features (privileges and restrictions). The number of features is the same for all nodes.

Such that: $V_{i,k}(0) = (v_{i,1}, v_{i,2}, v_{i,3}, ..., v_{i,n})^T$, and n is the number of characteristics of the node i.

### C. Intrusion Detection Approach

Figure 3 shows the functioning of our approach:

*1)* The node presents the data source.

*2)* Normal node profile is built based on the permissions and restrictions of the node. This construction of the profile is achieved by a unit of construction of normal profile.

*3)* A collection unit is responsible for capturing information (data), which allows to build the current behavior a node.

*4)* A unit of analysis and control is responsible for comparing the current behavior with the normal profile of a node. In case of deviation the behavior of a node from its normal profile, an action is sent to the node.

*5)* The action is the behavior of the IDS to a possible intrusion; an alarm is transmitted to inform the node and its neighbors of the intrusion and that correction of the problem is required.

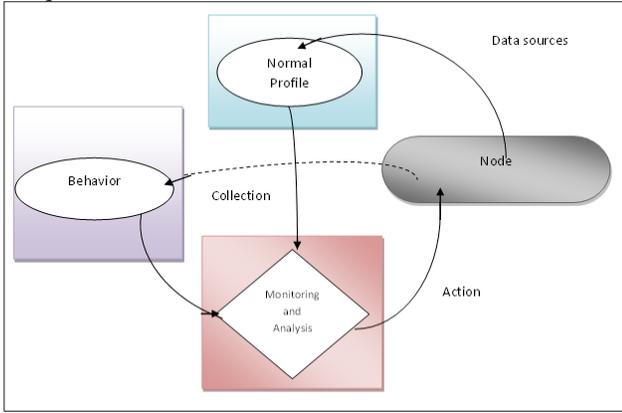

Figure 3. Step to the detection.

*1)* Normal profile.

Normal profile is constructed taking into account the activities of the user, as the preferred tool, work habits, typing speed, etc. In our approach the construction of normal profile is focused on: frequency; program CPU, I/O, other resources, and denied executions. We create profile based-on legitimate user data and represent a user's typical behavior.

User can have permissions and restrictions, which are representing by a vector of features. For example, a record (0110000010000000000110000100111010100000) is a vector of features (permissions and restrictions), where *1* is the permission, and *0* is the restriction.

*2)* Anomaly

All user-based anomaly detection schemes for intrusion detection are intrusive behaviors is, by its very nature, anomalous [13]. User is acting in an abnormal manner then the actions of that user can be classified as intrusive. Behaviors can be determined to be abnormal through a comparison against a user profile that represents a user's typical behavior.

Users establish a profile based-on the number and types of commands they employ. Thus, if a system can discern this profile, and commands are employed "outside" of this profile, then the system should flag this action as a potential intrusion. Commands are presented by authentication of user, which appear in the same sequence.

User profile can take on many forms, is based upon either an individual's behavior and/or the typical behavior of the individuals in a functional group. Any time that the system is not operating in a normal manner there is an increased likelihood that an intruder is (or was) present on the system. For example, a record:

(0010000010000110000110000100111010100010) is abnormal vector of features.

*D. Intrusion Detection*

The detection phase identifies abnormal vectors. At each time interval (t), each node collects a feature vector $V_{i,k}(t)$. This feature vector represents the behavior of a node in this time interval.

Detection of abnormal vector is based on the vector of normal profile $V_{i,k}(0)$ previously established.

To determine abnormalities, we calculate the distance between $V_{i,k}(0)$ and $V_{i,k}(t)$ using:

$$\sum_{\kappa=1}^{n} |Vi,\kappa(0), Vi,\kappa(t)| \qquad (1)$$

Such that n is the number of characteristics of the node *i*.

At each node, we calculate the distance of projection of each feature vector $V_{i,k}(t)$. We class $V_{i,k}(t)$ abnormal, if the projection distance calculated exceeds a predefined threshold:

$$\begin{cases} d(Vi,\kappa(0),Vi,\kappa(t)) = 0 : Normal \\ d(Vi,\kappa(0),Vi,\kappa(t)) \geq 0 : Anomaly \end{cases} \qquad (2)$$

*E. Application Example*

In a Smart Office, we have a set of nodes that have access to services and information.

The Smart Office has a central manager, whose management tasks are to access services and information in the network.

The number of nodes connected to the Smart Office is very large, which limits the work of central manager and encumbers management nodes.

Our idea is based on nodes and their behaviors. Each node accesses the services and information of Smart Office with network authentication. This authentication allows the authorizations and restrictions of services and information to nodes.

Generally, an intruder accesses the network via a node with the objective to access network services. Network protection depends on the protection of each node.

As a possible scenario to demonstrate the system design, we assume that Marc works in a Smart Office. He uses his PC to work and access network services and information requiring an authentication and access control mechanism. This authentication gives Marc the right to print, consult the database and send emails. Marc cannot update database, share data, or use the scanner. All tasks accomplished by Marc are tracked. Then his privileges are: print, consult the database and send emails, while his restrictions are: update database, share data and use scanner. The normal behavior of Marc is expressed by all these restrictions and privileges. This allows to build the vector features $V_{i,k}(0)$ with an initial time 0.

$V_{i,k}(0)$ = {print, consult, email, update, share, scan}
Such that I = name of the node (Marc), and k = number of privileges and restrictions (= 6).

*1) Intrusion Detection*

When connecting to the smart office, the profiling unit built the feature vector corresponding to the normal profile of Marc based on his privileges and restrictions.

In a time interval (t), the collection unit collects information about the current behavior of Marc, this allows to build the feature vector $V_{i,k}(t)$ corresponding to his current behavior. This vector shows the values of privileges and restrictions that Marc had accessed during this time interval.

The unit of analysis and detection calculates the distance between $V_{i,k}(t)$ and $V_{i,k}(0)$ to class the abnormal vectors (behavior):

If $d(Vi,\kappa(0),Vi,\kappa(t)) = 0$ : Normal behavior (3)
Else : Anomaly

Our idea allows checking all what comes from nodes, for monitoring the behavior of the user (node). The analysis and detection is performed in each node independently of the other nodes.

*2) Positioning of the solution*

IDSs shown in the related work are developed for sensor networks, which limits their generalization to ubiquitous networks. We also have the problem of overload sensors, and update database the profiles and signatures.

Our solution is based on the authentication by the permissions and restrictions of access to information and services on ubiquitous network. As such, there is no need to update database of the normal profile.

TABLE II. COMPARISON OF IDSS

| IDSs | Features | |
|---|---|---|
| | Construction of normal profile | Data source |
| SUIDS | Learning phase | Event Registration |
| Application-led | Learning phase | Data packets |
| Cluster-based | Learning phase | Network packets |
| ACP-based | Learning phase | Network packet |
| Our approach | Based on permissions and restrictions to access network. | Permissions and restrictions |

As the table shows, all works presented in the related work section are supported on the normal profile; require a learning phase for the construction of the normal profile. This phase limits the scalability of the solutions (support of new nodes). In our approach the construction of the normal profile does not require the learning phase, it based on permissions and restrictions to access network.

For the data source, our approach uses the permissions and restrictions of the node, and other approaches are based on networks or data packet.

In our solution, the detection is done on each node individually of its neighbors, limiting the cooperation in the detection. This is what we focus on. The design of schemes for achieving complete or partial cooperation is a topic of future research.

## IV. CONCLUSION AND FUTURE WORK

The principal objective of this work was to ensure the safety of ubiquitous networks. We developed an approach to detect intrusions; our approach allows to monitor the safety of nodes and the network. It consists in searching anomalies that could lead to possible attacks, and to take action against such attacks.

We then described our approach; we introduced a new way of constructing the normal profile of users based on authentication. This approach allows a more flexible analysis and detection, which avoids the update of the database of the normal profile. We interest to the behavior of the node to the network.

Our idea realizes a protection of nodes and assures security of network. This protection is proved by authentication abilities to build user profile.

The problem is still open, while the nodes act individually with their neighbors, limiting the cooperation in detection. In some case which attacks are complexes, the node will not be able to identify intrusion and intruder.

This article is a work in progress, presenting ideas to be developed in our future work.

As future work, we will complete the intrusion detection architecture, investigate the methods to analyze audit data for intrusion detection, formulate and evaluate our intrusion detection approach.